\documentstyle[psfig,12pt]{book}

     \makeatletter
     \font\scriptsize=cmr8
     \def\nothing#1{}
     \newdimen\earraycolsep
     \def\eqnarray{\let\@currentlabel\theequation
     \global\@eqnswtrue\m@th
     \global\@eqcnt\z@\tabskip\@centering\let \\\@eqncr
     $$\halign to\displaywidth\bgroup\@eqnsel\hskip\@centering
     $\displaystyle\tabskip\z@{##}$&\global\@eqcnt\@ne
     \hskip 2\earraycolsep \hfil$\displaystyle{##}$\hfil
     &\global\@eqcnt\tw@ \hskip 2\earraycolsep 
     $\displaystyle\tabskip\z@{##}$\hfil
     \tabskip\@centering&\llap{##}\tabskip\z@\cr}
     \renewcommand{\theequation}{\arabic{equation}}
     \renewcommand{\thetable}{\arabic{table}}
     \renewcommand{\thefigure}{\arabic{figure}}
     \def\title{\chapter}
     \renewcommand\chapter{\ifodd\c@page\clearpage\else\cleardoublepage\fi
                    \global\@topnum\z@
                    \@afterindenttrue
                    \secdef\@chapter\@schapter}
     \def\@makechapterhead#1{%
       \vspace*{120\p@}%
       {\parindent \z@ \raggedright
         {\bf #1}\par
         \nobreak
         \vskip 36\p@
       }}
     \def\author#1{{\pretolerance=10000 \raggedright \advance \leftskip by 
          1in \noindent #1 \vskip 1pc}}
     \def\affiliation#1{{\advance\leftskip by 1in \noindent #1 \vskip -1pc}}
     \def\refnote#1{{$^{\hbox{\scriptsize #1}}$}}

     \renewcommand\section{\@startsection{section}{1}{\z@}{2pc plus 1ex minus
         .2ex}{1pc plus .2ex}{\normalsize\bf}}
     \renewcommand\subsection{\@startsection{subsection}{2}{\z@}{1pc plus 1ex
         minus.2ex}{1pc plus .2ex}{\normalsize\bf}}
     \renewcommand\subsubsection{\@startsection{subsubsection}{3}{\parindent}
        {1pc plus 1ex minus.2ex}{-0.5em}{\normalsize\bf}}
     
     \def\AmS{{\protect\the\textfont2 
         A\kern-.1667em\lower.5ex\hbox{M}\kern-.125emS}}

     \def\p@LaTeX{{\family{times}\series{m}\shape{n}\selectfont 
         L\kern-.36em\raise.3ex\hbox{\scriptsize A}\kern-.15em 
         T\kern-.1667em\lower.7ex\hbox{E}\kern-.125emX}}
     
     \newlength{\colwidth}
     
     \def\@oddhead{\hfil}
     \def\@evenhead{\hfil}
     \def\@oddfoot{{\bf\hfil\thepage}}
     \def\@evenfoot{{\bf\thepage\hfil}}
     \def\fnum@figure{\footnotesize\raggedright{\bf Figure~\thefigure.}}
     \def\fnum@table{\normalsize\raggedright{\bf Table~\thetable.}}
     \long\def\@makecaption#1#2{\vskip 10\p@ {#1 #2\par}}
     \long\def\@makefntext#1{\setbox0=\hbox{$\m@th^{\@thefnmark}$}
          \noindent\hangindent=\wd0 \box0 #1}
     \def\centerfig#1#2#3#4{\vspace*{#2}\relax
         \centerline{\hbox to#1{\special{#4:#3.#4 x=#1, y=#2}\hfil}}}
     \newbox\@atbox
     \long\def\atable#1#2#3{\begin{table}[tbp]\centering\footnotesize
     \setbox\@atbox\hbox{#2}
     \parbox{\wd\@atbox}{\caption{#1}}\par\smallskip
     #2
     \par\smallskip\parbox{\wd\@atbox}{\raggedright #3}
     \end{table}}
     \def\@bibitem{\noindent \hangindent=2pc \hangafter=1}
     \def\thebibliography{%
     \section*{REFERENCES}%
     \bgroup\footnotesize
     \def\newblock{\hskip .11em plus.33em minus.07em}%
     \let\bibitem\@bibitem}
     \def\endthebibliography{\par\egroup}
     \def\@nbibitem#1{\noindent \hangindent=2pc \hangafter=1
     \refstepcounter{enumi}\hbox to 2pc{\arabic{enumi}.\hfil}%
     \immediate\write\@auxout{\string\bibcite{#1}{\arabic{enumi}}}}
     \def\numbibliography{%
     \section*{REFERENCES}%
     \bgroup\footnotesize
     \setcounter{enumi}{0}%
     \def\newblock{\hskip .11em plus.33em minus.07em}%
     \let\bibitem\@nbibitem}
     \def\endnumbibliography{\par\egroup}
     \makeatother
\begin{document}

\chapter{
SCALING LAWS, TRANSIENT TIMES AND SHELL EFFECTS IN HELIUM
INDUCED NUCLEAR FISSION
}

\author{Thorsten Rubehn, Kexing Jing, Luciano G. Moretto, Larry Phair,\\ 
Kin Tso, and Gordon J. Wozniak}

\affiliation{Nuclear Science Division,
Lawrence Berkeley National Laboratory,\\
University of California, 
Berkeley, California 94720, USA
}

\section{INTRODUCTION}
Fission excitation functions have been studied over the 
last decades and they have shown a dramatical
variation from nucleus to nucleus over the periodic table
\refnote{\cite{Rai67,Mor72,Kho66}}.
Some of these differences can be understood in terms of a changing
liquid-drop fission barrier, others are due to strong shell
effects which occur e.g. in the neighborhood of the double magic 
numbers $Z$=82 and $N$=126. 
Further effects may be associated with pairing and the angular 
momentum dependence of the fission barrier 
\refnote{\cite{Van73,Wag91,Hin91}}.
With the availability of newer accelerators, 
several studies have investigated 
heavy ion and high energy light particle induced 
fission \refnote{\cite{Wag91}}. These reactions show
a large deposit of energy, mass and most important angular
momentum. The strong dependence of 
the fission probability on the latter quantity makes 
comparisons to liquid drop model calculations difficult.  
The problem of extensive angular momentum, energy and mass 
transfer can be minimized by the use of light ion induced
fission at moderate bombarding energies.
In contrast to heavy ion reactions,
it has been shown that the fission barriers extracted
from low energy light ion induced fission reactions 
differ only slightly from liquid drop predictions
\refnote{\cite{Bec83,Mor95}}.

Fission rates have been successfully calculated on the base of
the transition state method introduced by
Wigner \refnote{\cite{Wig38}}, Bohr and Wheeler \refnote{\cite{Boh39}}.
Recent publications, however, claim
the failure of the transition state rates to account for the 
measured amounts of prescission neutrons or $\gamma$-rays in
relatively heavy fissioning systems 
\refnote{\cite{Hil92,Pau94,Tho93}}. 
This alleged failure has been attributed to the transient time
necessary for the so-called slow fission mode to attain its
stationary decay rate 
\refnote{\cite{Gra83a,Gra83b,Wei84,Gra86,Lu86,Lu90,Cha92,Fro93}}.
The experimental methods of these studies suffer from two
difficulties: First they require a possibly large correction
for post-saddle, but pre-scission emission; second, they
are indirect methods since they do not directly determine
the fission probability. 
Thus, the measured prescission particles can be emitted either
before the system reaches the saddle point, or during
the descent from saddle to scission. Only for the first
component deviations of the fission rate from its
transition state value would be expected. 
The experimental
separation of the two contributions, however, is fraught
with difficulties which make the evidence ambiguous. 
It seems therefore desirable 
to search for transient time effects by directly measuring
the fission probability and its energy dependence against
the predictions of the transient state method for a large
number of systems and over a broad energy range. 

In this paper, we show the results of a novel
analysis of fission excitation functions: 
The method allows the scaling of every single  excitation 
function of several compound nuclei produced in helium induced
reactions \refnote{\cite{Rai67,Kho66,Rub96}} exactly according to the 
transition state prediction onto a single straight line, once the 
shell effects are accounted for \refnote{\cite{Mor95}}. 
This analysis allows the investigation of transition state rates, 
shell effects, effective fission barriers and transient time effects 
directly from the data.

\begin{figure}[tb]
 %\vspace{15cm}
 \centerline{\psfig{file=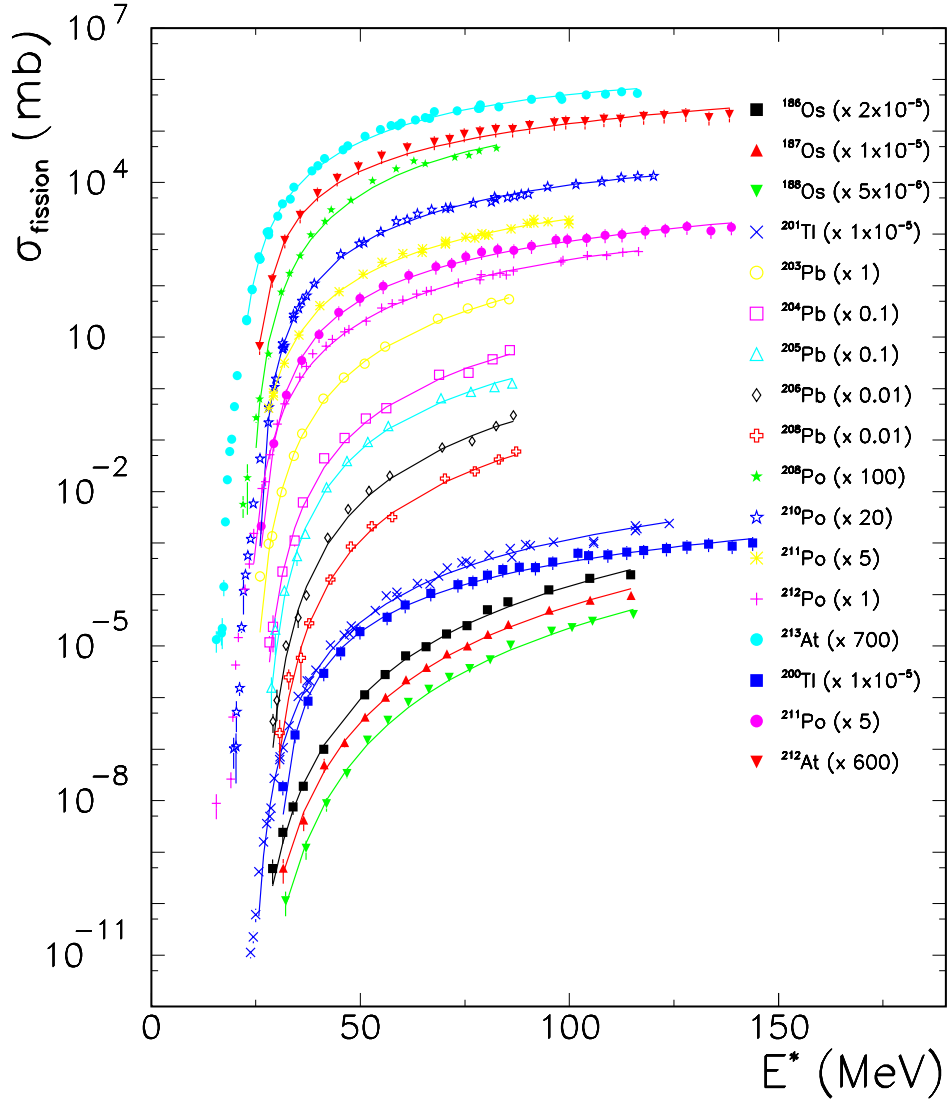,height=15cm}}
 \caption[]{Excitation function for fission of several compound
 nuclei formed in $^{3}$He and $^{4}$He induced reactions. The 
 different symbols correspond to the experimental data points,
 the solid line shows the results of a fit to the data using
 a level density parameter $a_{n} = A/8$.    
 The error bars denote the statistical and systematic errors
 combined in quadrature.
 }
 \label{exc_ftn}
\end{figure}

\section{ANALYSIS AND RESULTS}
\label{method}
The variety and accuracy of the measured fission excitation 
functions \refnote{\cite{Kho66,Rai67,Rub96}}, as 
shown in Fig.~\ref{exc_ftn}, enable us to search for
deviations from the predictions of the transition state
rates: In the following section we present a method 
that has been introduced in a recent letter\refnote{\cite{Mor95}}. 
It allows one to extract effective 
fission barriers and values for the shell effect that are 
independent of those obtained from the ground state masses.
Finally, a special way to plot the analysed data enables us to
investigate deviations from the transition state rates.

We start with a rather general 
transition state expression for the fission decay
width \refnote{\cite{Wig38,Boh39}},
\begin{equation}
 \Gamma_f \approx \frac{T_s}{2\pi} 
 \frac{\rho_s(E - B_f - E^s_r)}{\rho(E - E_r^{gs})}. 
\end{equation}
The latter allows one to write the fission cross section as follows: 
\begin{equation}
 \sigma_f = \sigma_0 \frac{\Gamma_f}{\Gamma_{total}}
 \approx \sigma_0 \frac{1}{\Gamma_{total}}
 \frac{T_s \rho_s (E - B_f - E^s_r)}{2\pi \rho_n (E - E_r^{gs})},
\end{equation}
where $\sigma_0$ is the compound nucleus formation cross section, 
$\Gamma_f$ and 
$\Gamma_n$ are the branching ratios for fission and neutron
emission, respectively, and $T_s$ is the energy dependent temperature
at the saddle; $\rho_s$ and $\rho_n$ are the saddle and ground
state level densities, $B_f$ is the fission barrier,
and $E$ the excitation energy. Finally, $E^s_r$
and $E^{gs}_r$ represent the saddle and ground state rotational
energies.

To further evaluate the expression, we use the form
$\rho(E) \propto \exp\big(2\sqrt{aE}\big)$ for the level density. 
This leads to:
\begin{equation}
 \log\Big( \frac{\sigma_f}{\sigma_0} \Gamma_T 
 \frac{2\pi \rho_n (E - E_r^{gs})}{T_s} \Big) =
 2 \sqrt{a_f (E - B_f - E_r^s)}.
 \label{scal}
\end{equation}
If the transition state null hypothesis holds, plotting the left
hand side of the equation versus $\sqrt{E - B_f - E_r^s}$ 
should result in a straight line. This equation has already
been used in Ref.~\refnote{\cite{Mor95a}} to show the scaling of all 
excitation functions obtained by the study of the emission of 
complex fragments from compound nuclei like 
$^{75}$Br, $^{90,94}$Mo, and $^{110,112}$In.
Since the neutron width dominates the total decay width in our
mass and excitation energy regime, we can write:
\begin{equation}
 \Gamma_{tot}  \approx \Gamma_n \approx K T_n^2 
 \frac{\rho_n(E - B_n - E_r^{gs})}{2\pi \rho_n(E - E_r^s}
\end{equation}
where $B_n$ represents the binding energy of the last neutron,
$T_n$ is the temperature after neutron emission, and 
$K = \frac{2 m_n R^2 g'}{\hbar^2}$ with the spin degeneracy
$g'=2$.

It is well known that the fission process is strongly influenced
by shell effects, which should be taken into account.
For the fission excitation functions
discussed in this paper the lowest excitation energies for
the residual nucleus after neutron emission are of the order of
15-20 MeV and therefore high enough to assume the asymptotic
form for the level density. Thus, an expression for that
quantity can be found:
\begin{equation}
 \rho_n(E-B_n-E_r^{gs}) \propto 
 \exp \big(2 \sqrt{a_n(E-B_n-E_r^{gs}-\Delta_{shell})} \big)
\label{rho_n}
\end{equation}
where $\Delta_{shell}$ is the ground state shell effect of the 
daughter nucleus ($Z,N-1$). 
For the level density at the saddle point we can use
\begin{equation}
 \rho_s(E-B_f-E_r^{s}) \propto 
 \exp \big(2 \sqrt{a_f(E-B_f^*-E_r^{s})} \big)
 \label{rho_s}
\end{equation}
since the saddle deformation implies small shell effects.
Deviations due to pairing, however, may be expected at very
low excitation energies. In Eq.~\ref{rho_s}, we introduced the
quantity $B_f^*$ which represents an effective fission barrier,
or, in other words, the unpaired saddle energy, i.e. 
$B_f^* = B_f + 1/2 g \Delta_0^2$ in the case of an even-even
nucleus and $B_f^* = B_f + 1/2 g \Delta_0^2 - \Delta_0$ for
nuclei with odd mass numbers. Here, $\Delta_0$ is the saddle
gap parameter and $g$ the density of doubly degenerate single
particle levels at the saddle.

\atable{Values of the effective fission barriers, $a_f/a_n$, 
  and shell effects. \label{res}}
  {
   \begin{tabular}{lccrr@{ $\pm$ }l}
    \noalign{\smallskip}\hline
   \multicolumn{1}{c}{Nuclide} &
   \multicolumn{1}{c}{Projectile} &
   \multicolumn{1}{c}{$B_f^*$ (MeV)} &
   \multicolumn{1}{c}{$a_f/a_n$} &
   \multicolumn{2}{c}{$\Delta_{shell}$ (MeV)}\\
    \noalign{\smallskip}\hline
    $^{213}$At & $^4$He & 20.1 & 1.036 & 9.7  & 1.5 \\
    $^{212}$At & $^3$He & 19.5 & 1.000 & 10.7 & 1.5 \\
    $^{212}$Po & $^4$He & 22.6 & 1.028 & 10.9 & 1.5 \\
    $^{211}$Po & $^4$He & 23.1 & 1.028 & 13.4 & 1.5 \\
    $^{211}$Po & $^3$He & 23.0 & 1.009 & 13.7 & 1.5 \\
    $^{210}$Po & $^4$He & 25.2 & 1.029 & 12.7 & 1.5 \\
    $^{208}$Po & $^4$He & 23.5 & 1.055 & 10.0 & 1.5 \\
    $^{208}$Pb & $^4$He & 27.1 & 1.000 & 10.2 & 2.0 \\
    $^{206}$Pb & $^4$He & 26.4 & 1.022 & 9.8  & 2.0 \\
    $^{205}$Pb & $^4$He & 26.4 & 1.001 & 11.8 & 2.0 \\
    $^{204}$Pb & $^4$He & 25.7 & 1.022 & 9.8  & 2.0 \\
    $^{203}$Pb & $^4$He & 24.1 & 1.021 & 10.0 & 2.0 \\
    $^{201}$Tl & $^4$He & 24.2 & 1.025 & 8.7  & 1.5 \\
    $^{200}$Tl & $^3$He & 25.1 & 0.995 & 12.1 & 1.5 \\
    $^{188}$Os & $^4$He & 23.2 & 1.025 & 1.4  & 2.0 \\
    $^{187}$Os & $^4$He & 22.7 & 1.022 & 3.2  & 2.0 \\
    $^{186}$Os & $^4$He & 22.4 & 1.020 & 1.5  & 2.0 \\
    \noalign{\smallskip}\hline
   \end{tabular}
  }

Finally, the usage of these expressions
allows us to study the scaling of the
fission probability as introduced in Eq.~\ref{scal}:
\begin{eqnarray}
 \frac{1}{2 \sqrt{a_n}} \log \Big(\frac{\sigma_f}{\sigma_0}
 \Gamma_{tot} \frac{2\pi\rho_n(E-E_r^{gs})}{T_s}\Big) =
%\nonumber\\
 \frac{\log R_f}{2\sqrt{a_n}}
 = \sqrt{\frac{a_f}{a_n}(E - B_f^* - E_r^s)}.
\label{rf_eq}
\end{eqnarray}
The values for $B_f^*$, $\Delta_{shell}$, and $a_f/a_n$ using
$a_n = A/8$ can be achieved by a three parameter fit of
the experimental fission excitation functions;
the best results of the fits are shown together with the 
experimental cross sections in Fig.~\ref{exc_ftn}. 
In order to make the figure more transparent, several
excitation functions have been multiplied by a factor 
which is indicated in the figure.
The formation cross sections $\sigma_0$ 
and the corresponding values for the maximum angular momentum
$l_{max}$ were taken from an optical model calculation. 
Finally, we computed the rotational energy at the saddle
assuming a configuration of two nearly touching spheres
separated by 2~fm. The results obtained from the fits
are also listed in Table~\ref{res}.

\begin{figure}[tb]
 %\vspace{10cm}
 \centerline{\psfig{file=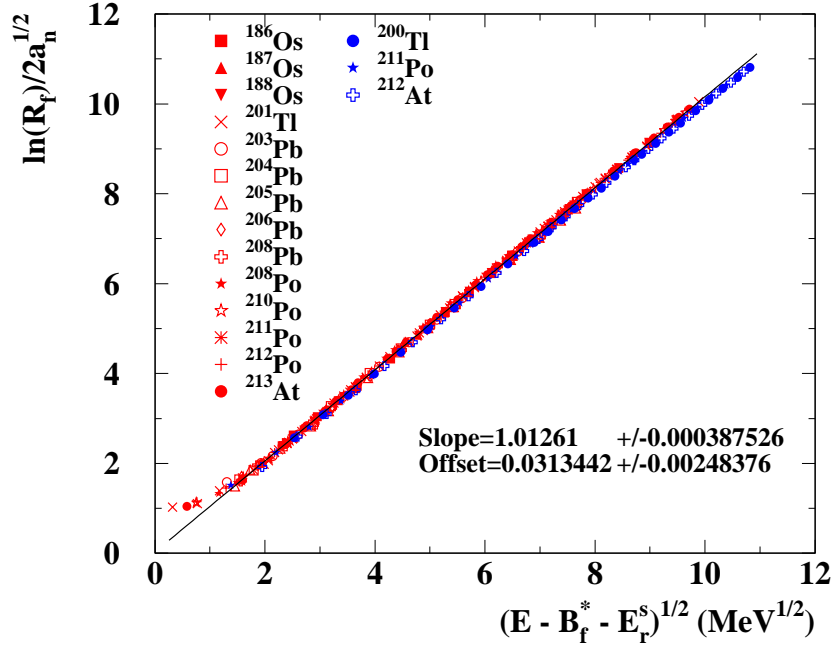,height=10cm}}
 \caption[]{The quantity $\frac{\log R_f}{2 \sqrt{a_n}}$ vs 
 the square root of the intrinsic excitation energy over the 
 saddle for fission of several compound nuclei as described
 in the text. The straight line represents a fit to the whole
 data set except for the lowest three points.
 }
 \label{rf}
\end{figure}

In Fig.~\ref{rf}, we now plot the left hand side of Eq.~\ref{rf_eq}
versus the square root of the effective excitation energy
above the barrier, $\sqrt{E-B_f-E_r^s}$, including the results
of the fits described above. 
A remarkable straight line can
be observed for all the investigated compound nuclei.
This scaling extends over six orders of magnitude in the
fission probability, although the shell effects are 
very strong for several nuclei.
Furthermore, a fit to the data results in a straight line 
that nearly goes through the origin and has 
a slope which represents the ratio $a_f/a_n$ very 
close to unity. 
The observed universal result and the lack of deviations
over the entire range of excitation energy  
indicates that the transition state null hypothesis 
and the above discussed equations for the level
density hold very well. 
The deviations from the straight line at very low
excitation energies are most likely due to slightly
different values of the level density at the saddle 
point from the Fermi gas values due to 
pairing effects. 

\begin{figure}[tbp]
 %\vspace{10cm}
 \centerline{\psfig{file=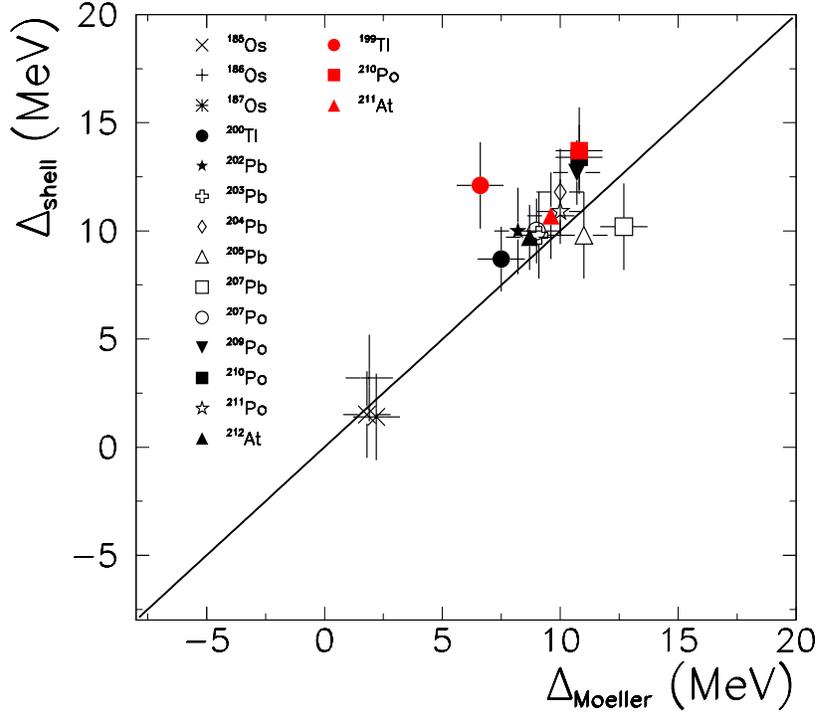,height=10cm}}
 \caption[]{Shell corrections for the daughter nuclei 
 ($Z,N-1$), extracted from fits to the excitation functions.
 The values of $\Delta_{shell}$ are plotted against the
 results determined from the ground state masses
 \protect\refnote{\cite{Mol94}}. 
 }
 \label{delta}
\end{figure}

As we have shown above, the employed method allows one to
extract values for the shell effect directly from the
data whereas the standard procedure determines
shell effects by the difference of 
the ground state mass and the corresponding liquid drop
value\refnote{\cite{Mye94}}. In Fig.~\ref{delta}, we show the resulting 
quantities of $\Delta_{shell}$ 
versus  a recent set of data\refnote{\cite{Mye94}} obtained by the 
standard method. A good correlation is observed,
especially if one reflects the difficulties connected with the 
liquid drop ground state baseline over the last 30 years.
We should point out that
the method shown above allows an independent 
determination of the shell effects, which is completely 
local since it only depends on the properties of the 
considered nucleus.

\begin{figure}[tbp]
 %\vspace{10cm}
 \centerline{\psfig{file=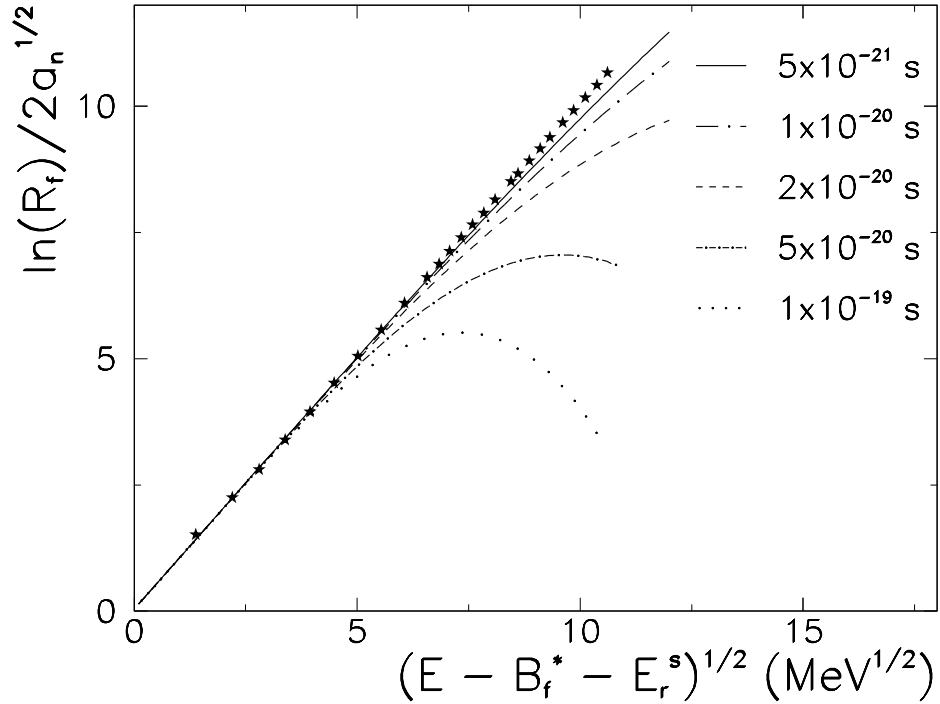,height=10cm}}
 \caption[]{Same as Fig.~\protect\ref{rf}. The lines
 represent calculations assuming that no fission occurs
 during a given delay time which is indicated in the
 figure. For further details see text.
 }
 \label{time}
\end{figure}

The presentations of the experimental data in Fig.~\ref{rf}
and Eq.~\ref{scal} imply the dominance of first chance fission. 
For the lower energies, calculations and experimental
investigations\refnote{\cite{Rub96}} verify this observation.
Even for the highest energy range
first chance fission still accounts for a large part
of the cross section but some uncertainties 
with the nuclear parameters, such as the barriers and
shell effects occur for the higher chance 
fissioning nuclei. 
However, it has been shown that the scaling still
holds very well even if only first chance fission
is investigated\refnote{\cite{Rub96}}.

Since our experimental results cover an excitation energy
range between 20 and 145 MeV, corresponding to life
times of the compound nucleus between 10$^{-18}$
and 10$^{-22}$ seconds, they should 
be sensitive to delay times in the first chance
fission probabilities.
In order to investigate this effect, we assume a step
function for the transient time effects. Then, the fission
width can be written as follows:
\begin{equation}
 \Gamma_f = \Gamma_f^{\infty} \int^{\infty}_{0} \lambda(t)
 \exp(\frac{-t}{\tau_{CN}}) dt = 
 \Gamma_f^{\infty} \exp(\frac{-\tau_D}{\tau_{CN}})
\end{equation}
where the quantity $\lambda(t)$ represents a step function
which jumps from 0 at times smaller
than the transient time $\tau_D$ to 1 for times larger than
$\tau_D$. Furthermore, $\Gamma_f^{\infty}$ denotes the
transition state fission decay width and $\tau_{CN}$ 
represents the life time of the compound nucleus. 
In Fig.~\ref{time}, we show the results of these 
calculations for the compound nuclei $^{211}$Po; 
the different lines indicate several 
transient times between 5$\times$10$^{-19}$ and 5$\times$10$^{-21}$
seconds. The shaded area indicates the uncertainty connected
with the contribution of first chance fission probability;
a detailed discussion on the latter can be found in
Ref.~\refnote{\cite{Rub96}}.
The calculated values show an obvious deviation
from the experimental data as long as the transient time
is longer than 10$^{-20}$ seconds. 
As already discussed in the introduction, this result is 
not in contradiction with recent measurements of prescission 
neutrons and $\gamma$ rays\refnote{\cite{Hil92,Pau94,Tho93}}, if these
particles are emitted during the descent  from saddle 
to scission.

\section{CONCLUSION}
\label{conclusion}

We have analysed and discussed fission excitation functions
according to a method which allows one to check 
the validity of the transition state rate predictions
over a large
range of excitation energies and a regime of compound nuclei masses
which are characterized by strong shell effects.  
Once these shell effects are accounted for, no deviation from
the transition state rates can be observed. 
Furthermore, the shell effects can be determined directly 
from the experimental data by using the above described procedure. 
In contrast to the standard method, there is no need to include 
liquid drop model calculations.
Finally, plotting the quantity $R_f$ allows one to search for 
an evidence of transient times as they have been discussed 
in a series of papers: Our results 
set an upper limit of 10$^{-20}$ seconds.

\bigskip
{\noindent\bf Acknowledgement}\\
This work was supported by the Director, Office of Energy Research,
Office of High Energy and Nuclear Physics, Nuclear Physics Division
of the US Department of Energy, under contract DE-AC03-76SF00098.

\end{document}